\documentclass[11pt,a4paper,parskip=half-]{scrartcl}
\usepackage{
  amsmath,
  amssymb,
  amsthm,
  amsfonts,
  graphicx,
  color,
  xcolor,
  colortbl,
  url,
  alltt,
  tikz,
  booktabs,
  tabularx,
  verbments,
  enumitem,
  multicol,
  textcomp,
  microtype,
  subfig,
}
\usepackage[hyperref=false]{scrhack}

\usepackage{libertine}
\usepackage[scaled=0.90]{helvet} 
\usepackage[T1]{fontenc}
\usepackage[utf8]{inputenc}

\usepackage[
  plainpages=false,
  colorlinks=true,
  breaklinks,
  pdfpagelabels,
]{hyperref}
\AtBeginDocument{\hypersetup{pdfauthor={\@author},pdftitle={\@title}}}

\usepackage[open,openlevel=0]{bookmark}
\usepackage[all]{hypcap}

\addtokomafont{subsection}{\normalsize\textsf\scshape}
\setkomafont{subsubsection}{\normalfont\sffamily\em}
\setkomafont{descriptionlabel}{\normalfont\sffamily\em}

\newcolumntype{x}[1]{>{\centering\hspace{0pt}}m{#1}}
\newcommand{\tn}{\tabularnewline}

\allowdisplaybreaks

\def\bold{\normalfont \bfseries}

\newcommand{\ben}{\begin{enumerate}}
\newcommand{\een}{\end{enumerate}}
\newcommand{\bit}{\begin{itemize}}
\newcommand{\eit}{\end{itemize}}
\newcommand{\bds}{\begin{description}}
\newcommand{\eds}{\end{description}}
\newcommand{\ttt}[1]{\texttt{#1}}

\newcommand{\fig}[1]{Figure~\ref{fig:#1}}
\newcommand{\subfig}[1]{\protect\subref{fig:#1}}
\newcommand{\tab}[1]{Table~\ref{tab:#1}}

\newcommand{\sect}[1]{\S\ref{sec:#1}}

\newtheorem{theorem}{Theorem}

\newcommand{\code}[1]{%
\begin{quote}%
\begin{alltt}%
#1%
\end{alltt}%
\end{quote}}

\usetikzlibrary{
  calc,
  fit,
  backgrounds,
  shapes,
  shapes.misc,
  positioning,
  patterns,
  decorations.pathreplacing
}

\tikzset{xcenter around/.style 2 args={execute at end picture={%
  \useasboundingbox let 
    \p0 = (current bounding box.south west), 
    \p1 = (current bounding box.north east),
    \p2 = (#1), 
    \p3 = (#2)
    in
    ({min(\x2 + \x3 - \x1,\x0)},\y0) 
    rectangle 
    ({max(\x3 + \x2 - \x0,\x1)},\y1);
}}}

\tikzset{
  shaded/.style=      {fill=gray!10},
  shadedLight/.style= {fill=gray!10},
  shadedMedium/.style={fill=gray!22},
  shadedDark/.style=  {fill=gray!35},
  faded/.style={opacity=0.5},
  outline/.style={rounded corners=0.8mm,color=gray!70},
}

\edef\defaultpgflinewidth{0.4pt}

\pdfpageattr{/Group <</S /Transparency /I true /CS /DeviceRGB>>} 

\def\copyright{
This work is licensed under the Creative Commons
Attribution 4.0 International License.\newline
\url{http://creativecommons.org/licenses/by/4.0/}}

\title{Emulating a large memory\\
with a collection of smaller ones\footnote{\copyright}}
\author{James Hanlon}
\date{June 2015}

\usepackage{scrpage2}
\hypersetup{
  pdfauthor={James Hanlon},
  pdftitle={Emulating a large memory with a collection of smaller ones}
}

\begin{document}

\maketitle

\begin{abstract}

Sequential computation is well understood but does not scale well with current
technology. Within the next decade, systems will contain large numbers of
processors with potentially thousands of processors per chip in order to deliver
continuing growth in performance.
Despite this, many computational problems exhibit little or no parallelism and
many existing formulations are sequential. It is therefore essential that
highly-parallel architectures can support sequential computation by emulating
large memories with collections of smaller ones, thus supporting efficient
execution of sequential programs or sequential components of parallel programs.
This paper demonstrates that a realistic parallel architecture with scalable
low-latency communications can execute large-memory sequential programs with a
factor of only 2 to 3 overhead, when compared to a conventional sequential
memory architecture.
This overhead seems an acceptable price to pay to be able to switch between
executing highly-parallel programs and sequential programs with large memory
requirements.
Efficient emulation of large memories could therefore facilitate a transition
from sequential machines by allowing existing programs to be compiled directly
to a highly-parallel architecture and then for their performance to be improved
by exploiting parallelism in memory accesses and the computation.
\end{abstract}

\section{Introduction}
\label{sec:introduction}

\def\draminfofootnote{
The central component of a DRAM is an \emph{array core}, a two-dimensional
array of cells and associated peripheral circuitry.  The array core is sized to
trade-off well between density and delay and energy per activation and refresh.
Array cores are limited to a modest size that grows very slowly with respect to
technology scaling due to their intrinsic capacitance.
For more details on the architecture of DRAM chips see Itoh~\cite{Itoh01} and
for general information on DRAM systems see Jacob, Spencer and
Wang~\cite{Jacob07}.}

The main component of a conventional sequential computer is a
monolithic memory that provides a uniform address space.  The memory is
typically implemented with a collection of DRAM arrays that are integrated in
one or more chips and connected with an interconnect specialised to transmit
control, data and address information, to provide efficient random
access.\footnote{\draminfofootnote}
The architecture of a DRAM system is therefore tightly coupled but inherently
distributed.  In this sense, a DRAM system is conceptually similar to a
distributed-memory parallel computer, which is composed of an array of
processors with local memories, connected by a communication network. A
parallel computer can thus be viewed as a more general system that provides
processing capability at each memory as well supporting arbitrary patterns of
communications between processors.

The similarities between conventional monolithic memory systems and the features
of a distributed parallel architecture lead to the question, and the subject of
this paper, \emph{how efficiently can a parallel computer emulate a conventional
memory system?}
The ability to do so efficiently would allow a parallel computer to support
sequential execution, in the same way that sequential computers are able
to efficiently emulate parallelism, an ability of considerable importance.
For parallel computers, as well as providing a means to execute legacy software
directly, efficient emulation of large memories would allow the degree of
parallelism to be matched with that of the problem being solved, by decreasing
the number of available processors and increasing the amount of memory
available to each one.

The critical factor in determining the ability of a distributed-memory
parallel computer to emulate a large memory is latency.
DRAM systems have been subject to around 40 years of optimisation in their
architecture and manufacturing processes to minimise access latency.
Latency in DRAM systems is therefore determined primarily by the intrinsic
physical characteristics of the component devices; specifically by the
transmission delay on wires, the dimensions of the component memory arrays and
the critical delay of control components.
The hypothesis of this paper however, is that because of the architectural and
physical similarities, the overheads for a more-general parallel computer
system are in fact relatively small and that it can therefore deliver an
efficient emulation with only a small constant-factor overhead in performance.

\subsection{Contributions and outline of the paper}

The contributions of this paper are:
%
%
(1) a proposal for a parallel architecture that can support sequential
programming techniques by emulating large memories;
%
(2) a model of the implementation of the architecture, based on current
production manufacturing technologies, that demonstrates its practicality and
scalability;
%
(3) parameters for the implementation model based on the state of the art;
%
and (4) a performance evaluation of the proposed architecture with results that
demonstrate that the efficiency of a memory emulation incurs just a small
constant-factor overhead.


The proposed parallel architecture is presented in \sect{architecture};
\sect{background} provides some background to the relevant technologies;
\sect{implementation} describes the implementation model;
\sect{parameters} presents the parameters and an analysis of the model's cost
and scaling;
\sect{methodology} describes the experimental methodology undertaken to
characterise the emulation performance;
\sect{results} presents the performance results;
\sect{related} discusses related work;
and \sect{conclusion} concludes.

\section{A general parallel architecture}
\label{sec:architecture}

\def\processorfootnote{
Microarchitectural optimisations to improve sequential performance are employed
at the expense of silicon area. The larger the processor, then fewer cores can
be integrated into a single chip. A tradeoff must be struck based on the
requirements of the processor.}

\def\dramfootnote{
A low cost-performance ratio is the primary driver in the production of
commodity DRAMs.  The basic memory cells and supporting circuits are replicated
millions of times on a chip and consequently the designs are highly optimised.
DRAM chips are produced in high volumes and are produced at a size which makes
the best trade-off between cost and performance. Consequently, their size has
remained virtually constant between technology generations.  Current DRAM chips
are manufactured with an area of around 100~mm$^2$ with a capacity of
approximately 0.3~Gbits~\cite[pp.~84-104]{ITRSExec12}.}

\def\foldedclosfootnote{
A Clos network is defined with three stages: input, middle and output, each
consisting of a number of nodes. Each of the middle-stage nodes is connected to
each of the input and output nodes.
The symmetry of a Clos network allows it to be \emph{folded} around the middle
stage and for the input and output networks to be merged using bidirectional
links~\cite[Ch.~3]{Jones97}.  This construction is similar to a fat
tree~\cite{Leiserson85}.
A folded Clos has two practical advantages: it makes the network more
convenient to package when connecting terminals that produce both input and
output, and messages can traverse shorter paths between terminals.}

A general-purpose parallel computer architecture is presented in this section
that is able to deliver an efficient emulation of a large memory.
Only the salient aspects relevant to the comparison are discussed; for more
details the reader is referred to~\cite[Ch.~5]{Hanlon14}.

The architecture consists of a collection of tiles, each containing a
processor, a memory and an interface to the network.
The tiles are connected by an interconnect that supports efficient
communication between all pairs of tiles.
The specific details of the processor architecture are not
important\footnote{\processorfootnote} except that the latency of moving data
into the network is minimised.
As well as mediating the transfer of data between the processor and network,
the network interface also supports direct memory access. This ability allows a
processor to read and write the memory of a different tile without involving
the processor of that tile, and thus incurring additional latency in the
transaction.
Conceptually, each tile can be viewed as a unit of processing, or a unit of
memory.

The structure of the interconnect is a folded Clos
topology~\cite[Ch.~3]{Jones97}, which provides a logarithmic diameter and
scalable bisection width. The former characteristic is essential to reduce
the distance a message travels between tiles and therefore its latency, and the
latter for maintaining low latency when there are many simultaneous
communications due to a parallel workload.
Folded Clos networks are part of a family of closely-related hypercubic
networks~\cite[Ch.~3]{Leighton91} but they have a particularly flexible
structure that is well suited to practical
implementations.\footnote{\foldedclosfootnote}
Specifically, folded Clos networks can be constructed from switches of any
fixed degree and their hierarchical structure allows a network to easily be
expanded or for the bisection width to be adjusted to meet the constraints of
an implementation technology.

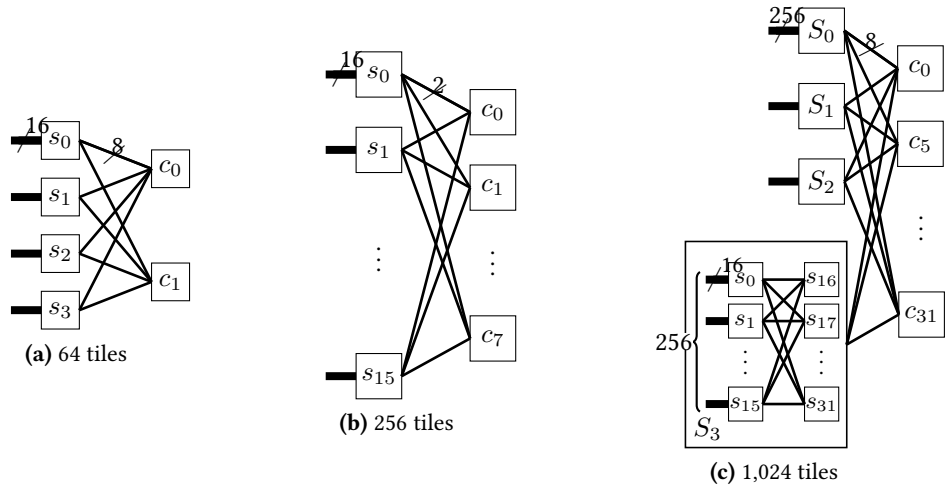
\begin{figure*}[t]
\centering
\subfloat[64 tiles]{
  
\def\switchHeight{0.5}
\def\switchSpacingY{0.75}
\def\switchSpacingX{1.2}
\newcommand{\switch}[1]{$s_{#1}$}
\newcommand{\coreswitch}[1]{$c_{#1}$}
\newcommand{\Switch}[1]{$S_{#1}$}

\begin{tikzpicture}[
baseline=($(e1)!0.5!(e2)$),
every node/.style={draw, inner sep=0pt, outer sep=0pt},
switch/.style={minimum height=\switchHeight cm, minimum width=\switchHeight cm},
link/.style={color=black,line width=1pt},
terminalLink/.style={color=black,fill,line width=3pt},
chip/.style={inner sep=2pt, line width=0.6pt},
]
\small

\draw node[switch] (e0) {\switch{0}}
	++(0,-\switchSpacingY) node[switch] (e1) {\switch{1}}
	++(0,-\switchSpacingY) node[switch] (e2) {\switch{2}}
	++(0,-\switchSpacingY) node[switch] (e3) {\switch{3}};

\draw ($(e0.east)!0.5!(e1.east)+(\switchSpacingX,0)$) node[switch] (c0) {\coreswitch{0}};
\draw ($(e2.east)!0.5!(e3.east)+(\switchSpacingX,0)$) node[switch] (c1) {\coreswitch{1}};

\draw[link] (c0.west) to node[draw=none,sloped,pos=0.65, label={[above=0mm,xshift=-0.5mm] 8}] {$\not$} (e0.east);
\foreach \a in {c0,c1}
\foreach \b in {e0,e1,e2,e3}
		\draw[link] (\a.west) -- (\b.east);

\draw[terminalLink] (e0.west) to node[draw=none,sloped,pos=1.2, label={[above=1mm,xshift=1mm] 16}] {$\not$}  ($(e0.west)+(-0.3,0)$);
\foreach \a in {e0,e1,e2,e3}
	\draw[terminalLink] (\a.west) -- ($(\a.west)+(-0.4,0)$);

\end{tikzpicture}%

  \label{fig:clos-64}}
\qquad
\subfloat[256 tiles]{
  
\def\switchHeight{0.6}
\def\switchSpacingY{1}
\def\switchSpacingX{1.2}
\newcommand{\switch}[1]{$s_{#1}$}
\newcommand{\coreswitch}[1]{$c_{#1}$}
\newcommand{\Switch}[1]{$S_{#1}$}

\begin{tikzpicture}[
baseline=($(e0)!0.5!(en)$),
every node/.style={draw, inner sep=0pt, outer sep=0pt},
switch/.style={minimum height=\switchHeight cm, minimum width=\switchHeight cm},
link/.style={color=black,line width=1pt},
terminalLink/.style={color=black,fill,line width=3pt},
chip/.style={inner sep=2pt, line width=0.6pt},
vdots/.style={draw=none},
]
\small


\draw node[switch] (e0) {\switch{0}}
	++(0,-\switchSpacingY) node[switch] (e1) {\switch{1}}
	++(0,-3*\switchSpacingY) node[switch] (en) {\switch{15}};
\draw ($(e1)!0.45!(en)$) node[vdots] {$\vdots$};

\draw ($(e0.east)!0.5!(e1.east)+(\switchSpacingX,0)$) node[switch] (c0) {\coreswitch{0}}
	++(0,-\switchSpacingY) node[switch] (c1) {\coreswitch{1}}
	++(0,-2*\switchSpacingY) node[switch] (cn) {\coreswitch{7}};
\draw ($(c1)!0.45!(cn)$) node[vdots] {$\vdots$};

\draw[link] (c0.west) to node[draw=none,sloped,pos=0.65, label={[above=0mm,xshift=-0.5mm] 2}] {$\not$} (e0.east);
\foreach \a in {c0,c1,cn}
\foreach \b in {e0,e1,en}
		\draw[link] (\a.west) -- (\b.east);

\draw[terminalLink] (e0.west) to node[draw=none,sloped,pos=1.2, label={[above=1mm,xshift=1mm] 16}] {$\not$}  ($(e0.west)+(-0.3,0)$);
\foreach \a in {e0,e1,en}
	\draw[terminalLink] (\a.west) -- ($(\a.west)+(-0.4,0)$);

\end{tikzpicture}%

  \label{fig:clos-256}}
\qquad
\subfloat[1,024 tiles]{
  
\def\switchHeight{0.6}
\def\lineWidth{0pt}
\def\switchSpacingY{1}
\def\switchSpacingX{1.4}
\newcommand{\switch}[1]{$s_{#1}$}
\newcommand{\coreswitch}[1]{$c_{#1}$}
\newcommand{\Switch}[1]{$S_{#1}$}

\begin{tikzpicture}[
baseline=($(e1)!0.5!(e3)$),
every node/.style={draw, inner sep=0pt, outer sep=0pt,},
switch/.style={minimum height=\switchHeight cm, minimum width=\switchHeight cm},
smallSwitch/.style={minimum size=0.45 cm},
link/.style={color=black,line width=1pt},
terminalLink/.style={color=black,fill,line width=3pt},
subClos/.style={inner sep=3pt, line width=0.6pt},
vdots/.style={above=-5,draw=none},
]
\small

\draw node[switch] (e0) {\Switch{0}}
	++(0,-\switchSpacingY) node[switch] (e1) {\Switch{1}}
	++(0,-\switchSpacingY) node[switch] (e2) {\Switch{2}}
	++(0,-1.8*\switchSpacingY) node[draw=none] (e3) {};

\begin{scope}[
switch/.style={minimum height=0.5*\switchHeight cm, minimum width=0.5*\switchHeight cm, fill=gray!20},
]
\def\scaling{0.5}

\coordinate (p) at ($(e3)+(0,0.5)$);
\foreach \x / \y / \z in {0/0/16,1/1/17,3/15/31} {
\draw ($(p)-(0,0.55*\x)$) node[smallSwitch] (subc\x) {\footnotesize$s_{\z}$};
\draw ($(p)-(1,0.55*\x)$) node[smallSwitch] (sube\x) {\footnotesize$s_{\y}$};
}
\draw ($(subc1)!0.4!(subc3)$) node[draw=none] {\footnotesize$\vdots$};
\draw ($(sube1)!0.4!(sube3)$) node[draw=none] {\footnotesize$\vdots$};

\foreach \x in {0,1,3} {
	\foreach \y in {0,1,3} {
		\draw[link] (subc\x.west) -- (sube\y.east);
	}
}

\draw[terminalLink] (sube0.west) to node[draw=none,sloped,pos=1.6, 
	label={[above=1mm,xshift=1.5mm] \footnotesize 16}] {$\not$}  ($(sube0.west)+(-0.2,0)$);
\foreach \a in {sube0,sube1,sube3}
	\draw[terminalLink] (\a.west) -- ($(\a.west)+(-0.3,0)$);

\draw ($(sube3.south)-(0.5,0.1)$) node[draw=none] (label) {\Switch{3}};

\coordinate (a) at ($(sube0)-(0.6,-0.4)$);
\coordinate (b) at ($(subc1)+(0,-0.5)$);
\node[subClos,fit=(a) (sube3) (subc0) (b) (label)] (e3) {};

\draw [thick, decoration={brace, mirror, raise=2mm}, decorate] 
($(sube0)-(0.4,-0.1)$) -- ($(sube3)-(0.4,0.1)$) node [pos=0.5,xshift=-3mm,left,align=right,draw=none,fill=white] {256}; 

\end{scope}

\draw ($(e0.east)!0.5!(e1.east)+(1,0)$) node[switch] (c0) {\coreswitch{0}}
	++ (0,-1) node[switch] (c1) {\coreswitch{5}};
\draw ($(e2.east)!0.5!(e3.south east)+(1,0)$) node[switch] (cn) {\coreswitch{31}};

\draw ($(c1)!0.45!(cn)$) node[draw=none] {$\vdots$};

\draw[link] (c0.west) to node[draw=none,sloped,pos=0.65, label={[above=0mm,xshift=-0.8mm] 8}] {$\not$} (e0.east);
\foreach \a in {c0,c1,cn}
\foreach \b in {e0,e1,e2,e3}
	\draw[link] (\a.west) -- (\b.east);

\draw[terminalLink] (e0.west) to node[draw=none,sloped,pos=1.2, label={[above=1mm,xshift=0mm] 256}] {$\not$}  ($(e0.west)+(-0.3,0)$);
\foreach \a in {e0,e1,e2}
	\draw[terminalLink] (\a.west) -- ($(\a.west)+(-0.4,0)$);

\end{tikzpicture}%

  \label{fig:clos-1024}}

\caption{Example 32$\times$32 crossbar switch topologies for various size
  folded Clos networks. In the 64- and 256-tile networks, edge switches
  (labelled $s_i$) connect 16 tiles and core switches (labelled $c_i$) use all
  32 links to connect to the previous stage.
  The 1,024-tile network is constructed by replicating a 256-tile network, with
  twice the number of core switches than \subfig{clos-256}, four times and
  connecting it with 32 core switches, making it three-stage.  This
  construction maintains capacity between stages and has a logarithmic diameter
(2 or 3 for the examples) in the number of tiles.}

\label{fig:clos-topologies}
\end{figure*}

It is beneficial to use high-degree switches to implement a folded Clos network
because, by connecting multiple terminals to each one, the network diameter can
be reduced.
A degree-32 switch provides a good balance between connectivity and
implementation complexity. With two stages it can connect up to 256 tiles,
which fits well into the target economical chip area
(see~\sect{chip-parameters}).
A degree-64 switch can connect up to 1,024 tiles with two stages but this size
of system far exceeds the target area.
Furthermore, with a degree-32 switch it is practical to use half the links to
connect tiles, leaving sufficient bandwidth in the set of remaining links.
\fig{clos-topologies} shows example folded Clos networks connecting 64, 256 and
1,024 tiles.

It is interesting to note that, in contrast with parallel programs, execution
of sequential programs will not induce any concurrent communication traffic in
the network, and unless additional processes are run in parallel, each message
will travel without contention.  When this is the case, the interconnect needs
only to provide low latency and high bandwidth, and therefore, a simple routing
scheme can be used where messages travel along shortest paths.

Following the production and packaging model of commodity
DRAM,\footnote{\dramfootnote} the proposed parallel architecture can be
implemented with tens or hundreds of tiles on a single chip that is sized to
provide the best economic trade-off between performance and system cost.
Each processing chip contains a complete sub-folded Clos network with
replicated processing tiles, switches and communication links.
The hierarchical structure of the folded Clos network allows multiple chips to
be combined to directly extend the network.
Each chip contributes additional switches to form a core switching stage
and communication links connect switches on different chips, in the same way
they are connected on-chip.
Thus, a particular number of chips are chosen to build a system to provide a
certain memory capacity or processing capability.

\subsection{Emulation scheme}
\label{sec:emulation-scheme}

Typically, the requirements of local storage, which includes the program,
constant values and the stack, are small and can fit into the local memory of a
tile.
The remaining global storage, which is used for statically-allocated data and
the heap for dynamically-allocated data, can be stored in an emulated memory.
The emulation is performed with a collection of tiles that are managed by a
controller process. The controller receives access requests over a contiguous
address range from a sequential client program and distributes them over the
tiles by sending direct memory access read and write messages to them.

In the sequential (client) program, accesses to the emulated memory are written
as communication sequences, where an instruction to load a value at an address
(\ttt{addr}) in memory into a destination operand (\ttt{dest}) has the abstract
correspondence:

\begin{center}
\begin{minipage}[b]{2.5cm}
\ttt{LOAD} dest, addr
\end{minipage}
\qquad$\rightarrow$\qquad
\begin{minipage}[b]{2.5cm}
\begin{tabular}{ll}
\ttt{SEND} $c$, \ttt{READ} & \\
\ttt{SEND} $c$, addr & \\
\ttt{RECV} $c$, dest & \\
\end{tabular}
\end{minipage}
\end{center}

\noindent In this, a \ttt{READ} request and an address are sent to the
controller, and then when the value has been retrieved, it is received and
assigned to the destination operand.
Similarly, a write has the correspondence:

\begin{center}
\begin{minipage}[b]{2.8cm}
\ttt{STORE} value, addr
\end{minipage}
\qquad$\rightarrow$\qquad
\begin{minipage}[b]{2.5cm}
\begin{tabular}{ll}
\ttt{SEND} $c$, \ttt{WRITE} & \\
\ttt{SEND} $c$, addr & \\
\ttt{SEND} $c$, value & \\
\end{tabular}
\end{minipage}
\end{center}

\noindent In both, $c$ is a reference to a channel of communication, by which
data can be conveyed to the memory controller process.

If the requirements of local storage exceeded the capacity of local memory,
then there are two potential resolutions.
Either, the program or stack could be split between processors, and control
would transfer from one processor to the other when using either portions of
the stack.
Alternatively, a second emulated memory could be used.  For the purposes of
this paper and without loss of generality, only the simple case is considered
where local data does not exceed the local memory capacity.

\section{Background}
\label{sec:background}

This section discusses background to the implementation model: the silicon
interposer technology that is used to connect multiple chips, the memory
technology, the characteristics of on-chip wires, and the processor and switch
components.

\subsection{Interposer}

\def\processtechfootnote{
The \emph{process technology} or \emph{process geometry} of a chip refers to
the minimum half pitch of contacted metal lines, as defined by the
ITRS~\cite[pp.~5-6]{ITRSExec12}, which is the minimum feature size of a
circuit.  This has historically been an indicator of integrated-circuit scaling
and it has been DRAM that has exhibited the smallest metal pitch.
Larger process geometries produce higher yield because the technology is better
developed and so it is practical to produce larger die sizes.}

The most challenging aspect of the implementation of the parallel architecture
is the density of connectivity between chips that is required to extend the
folded Clos network and maintain a good bisection bandwidth.
Conventional packaging technology poses significant challenges for
interconnecting chips that contain large numbers of processors.
A high-density ball grid array (BGA) package can have up to approximately 2,800
pins. The Xilinx XCVU440 package, for example, has 2,892 pins with 1,327
I/Os~\cite{XilinxPinouts15}.  It is typical that approximately 40\% of package
pins are used for ground and power~\cite[Tab.~ORTC-4]{ITRSExec12}. However, for
bidirectional byte-wide links, only 82 such links could be connected with the
XCVU440 package; significantly less than the level of on-chip connectivity.

Printed circuit boards (PCBs) offer a minimum wiring pitch of about 0.15~mm and
a high-density BGA package would require approximately six wiring layers to
route tracks to each of the pins~\cite{Pfeil07}. The resulting density of
wiring on the board is also low compared to on chip; 16 adjacent links, for
instance, would occupy 38.4~mm.
Even for modest numbers of processors, PCB technology does not provide
sufficient connectivity between chips containing hundreds of processors to
obtain a scalable bisection width. Thus, with PCB connections, the
communication properties on chip do not extend off chip, complicating the
programming model.

The only current production technology that provides high levels of
connectivity between chips are silicon interposers. (However, in the future,
other forms of 3D integration would apply naturally to package these processing
chips in dense stacks.)
Silicon interposers are becoming an established technology that can be used as
a substrate to interconnect multiple processing chips with high-density wiring.
In essence, they can be thought of as a high-density PCB.
The interposer can be manufactured using a larger process geometry than the
chips it connects to provide a good tradeoff between yield and wiring
density\footnote{\processtechfootnote}.

As an example, Xilinx use a 65~nm 776~mm$^2$ silicon interposer in their
Virtex-7 FPGA~\cite{Xilinx11} to connect four 28~nm FPGA `slices' as a single
larger device.  This was the first production device to do so, with the
principal reason to for using a silicon interposer being an economic one.
Compared to contemporary devices, 776~mm$^2$ is a very large chip size, but at
the mature 65~nm technology, production costs are relatively low and yields are
high.
Producing a number of identical smaller devices and connecting them with an
interposer to produce a single larger device provides a good overall trade-off
between cost and performance.
The same approach is taken with the implementation of the proposed architecture,
using the characteristics of the Xilinx interposer are used as a basis.

\subsection{Memory technology}

\def\memdensityfootnote{
Faster memories are by design less dense, with control logic overheads
amortised over a smaller number of bit cells.}

\def\edramfootnote{
As eDRAM technology has matured, the access latency has approached that of
SRAM~\cite{Anand11}, especially for large memories where the flight-time across
the array can dominate access latency and the improved density of eDRAM can
significantly reduce this.
Its main application has been to build large on-chip caches, replacing SRAM,
for example in the BlueGene/L~\cite{Iyer05} and Power7~\cite{Barth10} chips.}

The choice of tile memory technology and capacity depends on obtaining a good
trade-off between access latency, area and its integration with microprocessor
logic.
Ideally, the access time of a memory should be matched with the time to execute
one basic operation, but access time is generally inversely related to density,
therefore faster memories require more silicon
area\footnote{\memdensityfootnote}.

There are two dominant forms of fast random access memory that can be
integrated with CMOS logic: static RAM (SRAM) and embedded dynamic RAM (eDRAM).
SRAM can be integrated directly, has fast access latency and good random access
performance but it has a relatively low density since it uses six transistors
per bit.
eDRAM is integrated with CMOS logic by adding three to six additional steps to
the fabrication process and is more dense than SRAM (by a factor of 2 to 3)
since it uses a single transistor-capacitor pair to store a
bit~\cite{ITRSSystemDrivers12}.\footnote{\edramfootnote}

In contrast, contemporary commodity DRAM is produced on a manufacturing process
that is optimised for density and power. Volume devices also lag logic
technology by several nodes due to a better cost-performance tradeoff.
DRAM is typically packaged on PCB modules, called dual-inline memory modules
(DIMMs), and connected with wire traces on a parent PCB.

\subsection{Wires}

The logic and wiring resources provided by very-large-scale integration (VLSI)
devices are plentiful and they continue to scale, but chip utilisation in terms
of computational performance is limited by two factors: the I/O bandwidth of
packages (i.e.\ getting data in and out) and the performance of long wires
relative to the size of a chip in terms of delay and power~\cite{Ho01} (i.e.\
moving data around a chip).

There are techniques for implementing power-efficient global interconnections
such as low-swing signalling schemes~\cite{Ho03a} and optimisation of wire
width and spacing~\cite{Chun05}, but an unavoidable effect of the scaling of
VLSI devices is that wire delays exceed a single clock cycle.  This makes
modular parallel architectures that can tolerate multi-cycle latencies an
attractive way to build systems~\cite{Ho03b}, and adds weight to the proposed
approach.

Wire delay is related to the square of the length of the distance a signal is
transmitted. Long wires are therefore unattractive.
A conventional technique to reduce wire delay is to insert repeating elements
at regular intervals along a wire~\cite[Chap.~4]{Ho03b}. This reduces the total
delay to a multiple of the delay between repeaters, at optimal intervals, this
produces a linear relationship between delay and length. The cost of this is
the delay through the repeaters and the logic area and power required for them.

\subsection{Processor and switch components}

The performance model of the parallel architecture and memory emulation uses
the XMOS XCore processor architecture~\cite{May09XS1}, for which detailed
information is available,  and because it is well suited to be integrated into
the proposed architecture because it provides low-latency communications.
The execution of a send operation, to transfer data between one processor and
another, moves data in a single cycle from the processor's registers into a
message buffer. In the next cycle, the message can be transmitted on to the
network, towards its destination.
Additionally, the provision of hardware support for a number of processes with
dedicated registers and a simple scheduler avoids context-switch overheads.

The size of the switch component is based on the INMOS C104
switch~\cite[Ch.~3]{May93}. The C104 is of degree 32 and provides the required
capabilities to construct a folded Clos network, particularly with its
routing scheme and ability to aggregate links.

\section{An implementation model}
\label{sec:implementation}

A VLSI floorplan of the architecture is described in this section, with a level
of detail to produce approximate, but not unrealistic, estimates of area and
wire delays to characterise the parallel machine performance.
The floorplan is used to produce figures for the dimensions of chips and their
components and wire lengths. These figures are then combined and compared to
show the relative proportions of processors and memory to switches and
interconnect wiring.
Building on this model, the specific technology parameters are presented in
\sect{parameters} and an associated model of performance is presented in
\sect{performance-model}.

To provide a baseline for the evaluation, an equivalent construction with a 2D
mesh network is presented. 2D meshes are a popular choice for an interconnect
because they are simple to implement in two dimensions on a chip. 2D meshes are
not however suitable for interconnecting large numbers of processors because
their diameter and bisection bandwidth do not scale well with increasing
numbers of attached processors.

\subsection{VLSI model}

The following simplified VLSI model is used to produce a floorplan for the
chip. It is based on those in Mead and Conway~\cite{Mead80} and
Ullman~\cite{Ullman84}.

\subsubsection{Metal layers}

\bit

\item A chip consists of a number of metal layers. Metal layer 1 (M1) is used
for logic components and the remaining metal layers M2, M3, M4 etc.\ are used
  to route wires.

\item A particular metal layer contains wires orientated in one direction and
adjacent metal layers contain wires in perpendicular directions. This is to
reduce crosstalk, where a signal in one circuit affects the signal in another
circuit.

\eit

\subsubsection{Wires}



\bit

\item Wires carry signals in one direction and all wires are half
  shielded to further reduce crosstalk. This is where a ground wire is placed
  either side of a wire pair. Minimum signal wire density consequently
  decreases by $\frac{1}{3}$.

\item All wires have repeaters inserted to minimise latency.

\item Wires with multicycle delays are pipelined by inserting flip-flops.

\item For sets of wires, banks of repeaters and flip-flops are required. To
  simplify their insertion, interconnect wires are routed in dedicated
  channels.

\eit

\subsubsection{Chip}

\bit

\item The area of a chip is taken to be the smallest rectangle in which the
  complete circuit fits.

\item External connections are made to I/O pads with a fixed driver circuit
component. The driver deals with the higher capacitance of external wires.

\eit

\subsubsection{Further assumptions and limitations}

Only the following essential aspects of
the system are captured in the layout to avoid over complicating the model:
\bit

\item the only components included in the layout are the processor core,
  memory, switch, pad driver circuitry and I/O pads (internal component wiring
  is not considered) and these are assumed to have a square footprint;

\item only the routing of the inter-switch and switch to chip I/O wiring is
  considered since this is the most significant component of the system in
  terms of wire lengths and density;

\item the remaining communication links between switches and processors are not
  explicitly accounted for in the area; it is assumed they can be routed over
  other resources, in the area not occupied by the interconnect wiring
  channels;

\item links are connected between components and it is assumed they can be
  connected from any position within the footprint of a component;

\item layout relating to wiring and I/O for power and clock signals is not
  considered, but provision is made for them in the available metal layers and
  chip I/Os.

\eit

\subsection{Folded-Clos layout}

A single chip contains a collection of tiles interconnected with a complete
sub-folded Clos network. The chip presents a number of external links from its
stage-two switches so that multiple chips can be connected with the links
between switches in different chips to directly extend the network.
Banks of additional switches are included to contribute to the third core stage
of a larger multi-chip network (note that the number of switches per chip is
constant with respect to the overall network size).
The layout of the folded Clos network has a recursive structure and is based on
an H-tree, an efficient embedding of a binary tree onto a 2D
surface~\cite[Ch.~8]{Mead80}.
An example layout for a two-stage 256-tile folded-Clos network relating to the
topology of one of the four sub-folded Clos networks in \fig{clos-1024} is
given in \fig{clos-layout}.

The core switches of the chip's folded Clos network are placed centrally and
half of the links are routed to the off-chip I/O pads.  The next stage of
switches is divided into four groups and each are placed in the centre of a
quadrant surrounding the core switches.
Connections are made from all of the core switches to each of the next stage
switches. This layout continues recursively in each quadrant for each
additional stage.  The minimum separation of groups of tiles is constrained by
the placement of switches and the width of the wiring channels between them.
All links from the contributed stage-three core switches are routed to off-chip
I/O pads, even though they will connect to the chip's second-stage switches,
because the connection pattern depends on the size of the multi-chip network.

Groups of switches are arranged in staggered sets to minimise the resulting
size of their bounding box. Their placement relative to one another is
constrained in two dimensions by the pitch of wires connecting horizontally and
vertically.  These are routed on the minimum of two metal layers, but
additional layers can be used to reduce the dimensions of a group.
Each group of switches has branching connections in both directions. This
connectivity restricts the horizontal and vertical placement of switches. A
switch arrangement is chosen to minimise the width of the group, subject to not
exceeding the height of its quadrant, since it can extend into the wiring
channel.
An example of the wiring pattern for a switch group is illustrated in
\fig{clos-layout}.
The total area of the interconnect is calculated as the sum of the area of the
wiring channels and all of the switch groups.

The pads and driver circuitry for off-chip I/O are positioned along one edge of
the chip due to the wiring pattern on the interposer.
A chip that contains $N$ tiles requires I/O for $2N$ links to
extend the network, $N$ from the core switches and $N$ from the contributed
bank of system core switches.

\subsection{2D-mesh layout}

\begin{figure*}
\centering
\subfloat[Clos layout]{
  
\input{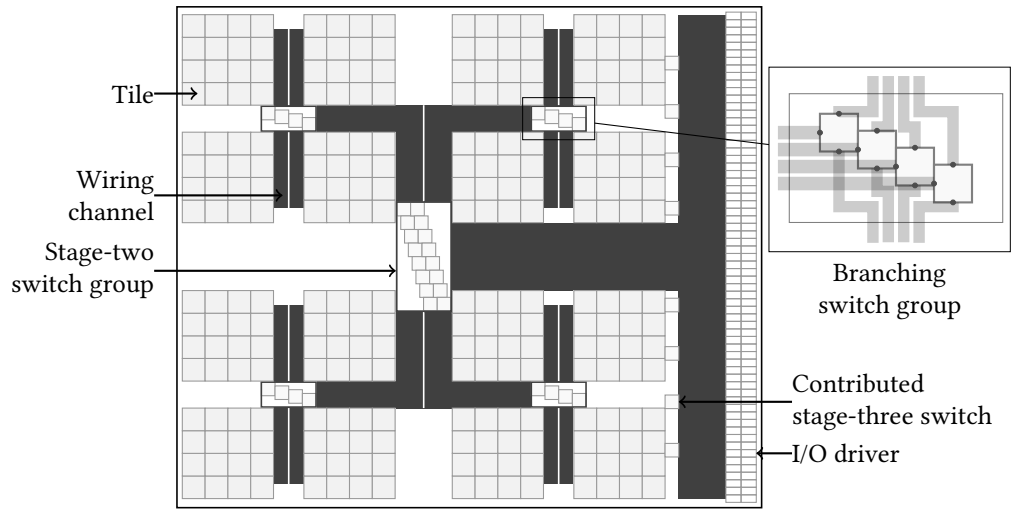}
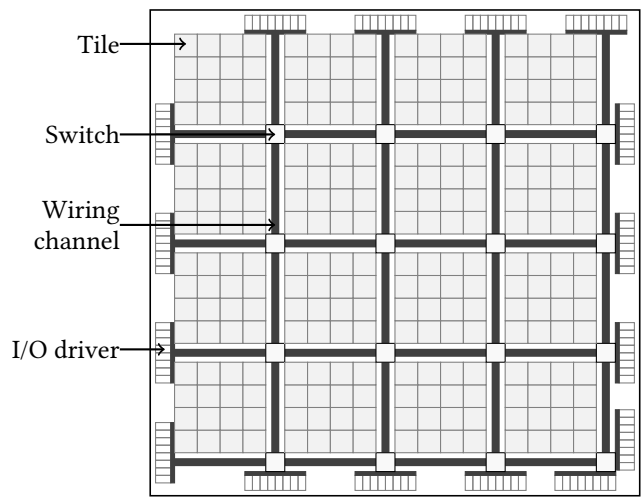%

  \label{fig:clos-layout}}

\bigskip
\subfloat[mesh layout]{
  
\def\procHeight{0.3}
\def\switchHeight{0.25}
\def\lineWidth{\defaultpgflinewidth}

\begin{tikzpicture}[
	every node/.style={draw=gray, inner sep=0pt, outer sep=0pt, line width=\lineWidth},
	switch/.style={minimum height=\switchHeight cm, minimum width=\switchHeight cm, anchor=north west,color=black,fill=gray!5},
	processor/.style={minimum height=\procHeight cm, minimum width=\procHeight cm, anchor=south west,shaded},
	links/.style={color=black!75,fill, line width=3},
	switchGroup/.style={inner sep=0pt,outer sep=\lineWidth, line width=0.6pt, color=black!75},
	separate/.style={color=white, line width=0.6pt},
	chip/.style={inner sep=2pt, line width=0.6pt, color=black},
	arrow/.style={line width=0.8pt},
	xcenter around={chip-box.south west}{chip-box.north east},
]
\small

\newcommand{\procGroup}[4]{
	\begin{scope}
	\node[processor,anchor=#2,minimum height=4*\procHeight cm,
		minimum width=4*\procHeight cm,draw=none] (proc-group-#3-#4) at ($#1$) {};
	\draw (proc-group-#3-#4.south east) node[switch, anchor=north west] (switch-#3-#4) {};
	\coordinate (base) at (proc-group-#3-#4.south west);
	\foreach \x in {0,1,2,3}
		\foreach \y in {0,1,2,3}
			\draw ($(base)+(\x*\procHeight,\y*\procHeight)$) node[processor] {};
	\end{scope}
}

\foreach \x in {0,1,2,3} {
	\foreach \y in {0,1,2,3} {
		\pgfmathtruncatemacro\result{\x+ (\y*3)}
		\pgfmathparse{(4*\procHeight)+\switchHeight}
		\pgfmathsetmacro\height{\pgfmathresult}
		\procGroup{(\x*\height,\y*\height)}{south west}{\x}{\y}
	}
}

\def\padwidth{1mm}
\def\iowidth{8mm}
\def\numpads{8}
\pgfmathsetmacro\last{\numpads-1}
\begin{scope}[on background layer,
	hpad/.style={draw,minimum width=\padwidth, minimum height=2mm},
	vpad/.style={draw,minimum height=\padwidth, minimum width=2mm},
	padwires/.style={color=black!75,fill},
]

\newcommand\tpads[4]{
	\foreach \i in {0,...,\last} {
		\node[hpad,yshift=0.5mm,#1] (pad-\i) at ($(#2)-(0.5*\iowidth,0)+(\i*\padwidth,0)$) {}; }
	\node[draw=none, fit=(pad-0) (pad-\last)] (#3-pads-#4) {};
	\draw[padwires] (#3-pads-#4.south west) rectangle ($(#3-pads-#4.south east)-(0,0.5mm)$);
}

\newcommand\bpads[4]{
	\foreach \i in {0,...,\last} {
		\node[hpad,yshift=-0.5mm,#1] (pad-\i) at ($(#2)-(0.5*\iowidth,0)+(\i*\padwidth,0)$) {}; }
	\node[draw=none, fit=(pad-0) (pad-\last)] (#3-pads-#4) {};
	\draw[padwires] (#3-pads-#4.north west) rectangle ($(#3-pads-#4.north east)+(0,0.5mm)$);
}

\newcommand\lpads[4]{
	\foreach \i in {0,...,\last} {
		\node[vpad,xshift=-0.5mm,#1] (pad-\i) at ($(#2)-(0,0.5*\iowidth)+(0,\i*\padwidth)$) {}; }
	\node[draw=none, fit=(pad-0) (pad-\last)] (#3-pads-#4) {};
	\draw[padwires] (pad-0.south east) rectangle ($(pad-7.north east)+(0.5mm,0)$);	
}

\newcommand\rpads[4]{
	\foreach \i in {0,...,\last} {
		\node[vpad,xshift=0.5mm,#1] (pad-\i) at ($(#2)-(0,0.5*\iowidth)+(0,\i*\padwidth)$) {}; }
	\node[draw=none, fit=(pad-0) (pad-\last)] (#3-pads-#4) {};
	\draw[padwires] (pad-0.south west) rectangle ($(pad-7.north west)-(0.5mm,0)$);
}

\coordinate (x) at ($(proc-group-0-3.north east)!0.5!(proc-group-1-3.north west)$);
\tpads{anchor=south west}{x}{top}{1}
\coordinate (x) at ($(proc-group-1-3.north east)!0.5!(proc-group-2-3.north west)$);
\tpads{anchor=south west}{x}{top}{2}
\coordinate (x) at ($(proc-group-2-3.north east)!0.5!(proc-group-3-3.north west)$);
\tpads{anchor=south west}{x}{top}{3}
\coordinate (x) at ($(proc-group-3-3.north east)$);
\tpads{anchor=south west}{x}{top}{4}
\node[draw=none, fit=(top-pads-1) (top-pads-4)] (top-pads) {};

\coordinate (x) at ($(switch-0-0.south)$);
\bpads{anchor=north west}{x}{bottom}{1}
\coordinate (x) at ($(switch-1-0.south)$);
\bpads{anchor=north west}{x}{bottom}{2}
\coordinate (x) at ($(switch-2-0.south)$);
\bpads{anchor=north west}{x}{bottom}{3}
\coordinate (x) at ($(switch-3-0.south)$);
\bpads{anchor=north east,xshift=-1.7mm}{x}{bottom}{4}
\node[draw=none, fit=(bottom-pads-1) (bottom-pads-4)] (bottom-pads) {};

\coordinate (x) at ($(proc-group-0-3.south west)!0.5!(proc-group-0-2.north west)$);
\lpads{anchor=south east}{x}{left}{1}
\coordinate (x) at ($(proc-group-0-2.south west)!0.5!(proc-group-0-1.north west)$);
\lpads{anchor=south east}{x}{left}{2}
\coordinate (x) at ($(proc-group-0-1.south west)!0.5!(proc-group-0-0.north west)$);
\lpads{anchor=south east}{x}{left}{3}
\coordinate (x) at ($(proc-group-0-0.south west)$);
\lpads{anchor=south east}{x}{left}{4}
\node[draw=none, fit=(left-pads-1) (left-pads-4)] (left-pads) {};

\coordinate (x) at ($(switch-3-3.east)$);
\rpads{anchor=south west}{x}{right}{1}
\coordinate (x) at ($(switch-3-2.east)$);
\rpads{anchor=south west}{x}{right}{2}
\coordinate (x) at ($(switch-3-1.east)$);
\rpads{anchor=south west}{x}{right}{3}
\coordinate (x) at ($(switch-3-0.north east)$);
\rpads{anchor=south west,yshift=1.7mm}{x}{right}{4}
\node[draw=none, fit=(right-pads-1) (right-pads-4)] (right-pads) {};

\end{scope}

\begin{scope}[on background layer]
\foreach \y in {0,1,2,3}
	\foreach \x / \xn in {0/1,1/2,2/3}
		\draw [links] (switch-\x-\y) -- (switch-\xn-\y);

\draw [links] (switch-0-3) -- (switch-0-3 -| left-pads-1.east);
\draw [links] (switch-0-2) -- (switch-0-2 -| left-pads-2.east);
\draw [links] (switch-0-1) -- (switch-0-1 -| left-pads-3.east);
\draw [links] (switch-0-0) -- (switch-0-0 -| left-pads-4.east);

\foreach \y / \yn in {0/1,1/2,2/3}
	\foreach \x / \xn in {0,1,2,3}
		\draw [links] (switch-\x-\y) -- (switch-\x-\yn);

\draw [links] (switch-0-3) -- (switch-0-3 |- top-pads-1.south);
\draw [links] (switch-1-3) -- (switch-1-3 |- top-pads-2.south);
\draw [links] (switch-2-3) -- (switch-2-3 |- top-pads-3.south);
\draw [links] (switch-3-3) -- (switch-3-3 |- top-pads-4.south);
\end{scope}

\node[chip,fit=(proc-group-0-0) (switch-3-0) (proc-group-0-3) (switch-3-3) (top-pads) (bottom-pads) (left-pads) (right-pads)] (chip-box) {};

\coordinate (p) at ($(chip-box.north west)-(0.4,0)$);

\coordinate (a) at ($(p)-(0,0)$);
\node[draw=none,anchor=east] (tileLabel) at ($(p)-(0,0.45)$) {Tile};
\draw[->,arrow] (tileLabel) -- ($(tileLabel)+(1.15,0)$);

\node[draw=none,anchor=east] (switchGroupLabel) at ($(p)-(0,1.65)$) {Switch};
\draw[->,arrow] (switchGroupLabel) -- (switchGroupLabel -| switch-0-3);

\node[draw=none,anchor=east,align=right] (channelLabel) at ($(p)-(0,2.85)$) {Wiring\\channel};
\coordinate (q) at ($(switch-0-2)+(0,0)$);
\draw[->,arrow] (channelLabel) -- (channelLabel -| q);

\coordinate (a) at ($(p)-(0,0)$);
\node[draw=none,anchor=east] (tileLabel) at ($(p)-(0,4.5)$) {I/O driver};
\draw[->,arrow] (tileLabel) -- ($(tileLabel)+(1.3,0)$);

\end{tikzpicture}%

  \label{fig:mesh-layout}}

\caption{Block diagrams of the 256-tile folded Clos and 2D mesh network layouts
for the processing chip.
In \subfig{clos-layout}, the topology corresponds to one of the four sub-folded
Clos networks in~\fig{clos-1024}.  The layout is based on an H-tree organisation
with groups of switches at each node organised in staggered sets to minimise
area subject to the constraints of their vertical and horizontal connections.
The driver circuitry and pad for each chip I/O is situated along the right-hand
side of the chip.
In \subfig{mesh-layout}, running around the edge of the chip is the driver
circuitry for each of the chip I/O pads.
Note that the connections from switches to processors and I/Os for power and
ground are not included in either layout.}

\end{figure*}

A 2D mesh network is laid out as an array of blocks of tiles where each
block connects to a single switch. The blocks are separated by wiring
channels accommodating the width and height of a switch, which is placed at the
corner of its block. Adjacent horizontal and vertical links connect directly
between switches.
The layout for a 256-tile 2D mesh network is given in \fig{mesh-layout}.

The pads and driver circuitry for the off-chip I/O are positioned at the
edges of the chip so that the mesh can be extended directly between adjacent
chips.
A chip that contains $N$ tiles requires I/O for $4\sqrt{N}-4$ links to extend
the network.

\subsection{Silicon interposer}
\label{sec:silicon-interposer}

The processing chips are mounted on a silicon interposer using a flip-chip
assembly.  This process bonds the top side of the chip with an array of solder
microbumps.  The interposer provides wiring traces between the microbumps to
connect links between switches on different chips.
The interposer is connected to a package substrate with through-silicon vias
(TSVs) that provide a connection through the interposer's substrate to
controlled-collapse chip-connection (C4) flip-chip bumps on the package
substrate.  These connections only bridge the power, ground and clocking I/Os.
The package substrate carries these connections out of the package, to the BGA
balls.

To extend the folded Clos topology, a set of chips are arranged in two rows,
either side of a wiring channel on the interposer, orientated with their wiring
area closest to this.
Each wire in the channel is used to connect between pads on two chips with
additional horizontal wires. The wiring channel contains sufficient wires so
that every pair of connections between the chips can be made.
The maximum height of the channel between two chips is therefore twice the
total pitch of the wires connecting a chip.
To extend the 2D mesh topology, connections are provided on the interposer
between the links of adjacent chips, directly extending the mesh.
In both networks, wiring for power, ground and clock signals is not considered
since provision is made in the available metal layers, but the area requirement
of the additional I/O pads required is accounted for directly.
\fig{interposer-layouts} illustrates the packaging of a set of processing chips
stacked on the silicon interposer for both networks.

\begin{figure*}
\centering

\def\chipWidth{2cm}
\usetikzlibrary{spy}
\begin{tikzpicture}[
	baseline=(package-substrate.base),
	chip/.style={minimum height=0.4cm, minimum width=\chipWidth, draw},
	interposer/.style={minimum height=0.3cm, minimum width=4.6cm, draw},
	interposerSub/.style={fill=black!50,draw},
	chipSub/.style={fill=black!50,draw},
	packageSubstrate/.style={minimum height=0.8cm, minimum width=6cm, draw, shadedMedium},
	tsv/.style={color=black, fill, line width=0.05cm},
	bgaBalls/.style={color=gray!90, fill},
	microBump/.style={color=black!60,fill},
	flipBump/.style={color=black!60,fill},
	package/.style={draw, rounded corners, line width=0.5pt},
	arrow/.style={line width=0.8pt},
	xcenter around={package-substrate.south west}{package-substrate.north east},
]
\small

\begin{scope}[spy using outlines={lens={scale=2.2}, connect spies}]

\draw (2cm,0) node[chip,anchor=east] (chip1) {}
	++(-\chipWidth-2,0) node[chip,anchor=east] (chip2) {};


\foreach \x in {1,2} {
	\draw[chipSub] ($(chip\x.north west)-(0,0.1cm)$) rectangle (chip\x.north east);
}

\draw node[interposer] (interposer) at (0,-0.44cm) {};
\draw[interposerSub] ($(interposer.south west)+(0,0.1cm)$) rectangle (interposer.south east);

\draw  (0,-1.1)  node[packageSubstrate] (package-substrate){package substrate};

\foreach \x in {1,2} {
	\foreach \y in {1,2,3,4,5,6,7,8} {
		\pgfmathsetmacro\result{\y/9}
		\draw[microBump] ($(chip\x.south west)!\result!(chip\x.south east)-(0,0.04cm)$) circle (0.04cm) node (ubump\x-\y) {};
    }
}

\foreach \x in {1,2,3,4,5,6,7,8,9,10,11,12,13,14,15,16} {
	\pgfmathsetmacro\result{\x/17}
	\coordinate (p) at  ($(interposer.south west)!\result!(interposer.south east)+(0,0.1)$);
	\coordinate (q) at ($(interposer.south)$);
	\draw[tsv] (p) -- (p |- q) {};
	\coordinate (tsv\x) at ($(p)!0.5!(p |- q)$);
	\draw[flipBump] ($(p |- q)-(0,0.05cm)$) circle (0.05cm);
}

\foreach \x in {1,2,3,4,5,6,7,8,9,10,11,12} {
	\pgfmathsetmacro\result{\x/13}
	\draw[bgaBalls] ($(package-substrate.south west)!\result!(package-substrate.south east)-(0,0.125cm)$) circle (0.12cm) node (ball\x) {};
}

\spy [black, minimum width=2.8cm, minimum height=2cm] on (-1.2,-0.45) in node (zoom) [right] at (-3,1.5);
\end{scope}

\node[anchor=west] (tsvLabel) at ($(zoom.west)+(3.6,-0.22)$) {Through-silicon via};
\draw[arrow,->] (tsvLabel.west) -- ($(tsvLabel.west)-(1,0)$);

\node[anchor=west] (ubumpLabel) at ($(zoom.west)+(3.6,0.45)$) {Microbump};
\draw[arrow,->] (ubumpLabel.west) -- ($(ubumpLabel.west)-(1.1,0)$);

\node[anchor=west] (bumpLabel) at ($(zoom.west)+(3.6,-0.7)$) {C4 bump};
\draw[arrow,->] (bumpLabel.west) -- ($(bumpLabel.west)-(1,-0.26)$);

\coordinate (p) at ($(chip1.east)+(1.25,-0.1)$);

\node[anchor=west] (chipLabel) at ($(p)+(0,0.1)$) {Processing chip};
\draw[arrow,->] (chipLabel.west) -- ($(chipLabel.west)-(1.25,0)$);

\node[anchor=west] (interposerLabel) at ($(p)+(0,-0.3)$) {Interposer};
\draw[arrow,->] (interposerLabel.west) -- ($(interposerLabel.west)-(1.05,0)$);

\node[anchor=west] (bgaLabel) at ($(p)-(0,1.53)$) {BGA ball};
\draw[arrow,->] (bgaLabel.west) -- ($(bgaLabel.west)-(0.6,0)$);


\end{tikzpicture}%

\caption{Cut-through view of a multi-chip package with a silicon interposer.
  A collection of chips are stacked using a flip-chip assembly onto solder
  microbumps, connecting them to the silicon interposer. The interposer
  provides connectivity between the chips, essentially as a high-density PCB.
  TSVs provide connectivity though the interposer's own substrate and these are
connected with C4 solder bumps to the package substrate.}

\label{fig:interposer-section}

\bigskip
\subfloat[Clos layout]{
  
\def\chipsize{18mm}
\def\wirewidth{16mm}

\begin{tikzpicture}[
	chip/.style={draw,minimum size=\chipsize, fill=white,shaded},
	io/.style={draw,minimum height=\chipsize,minimum width=3mm,shadedDark},
	package/.style={, draw, inner sep=2mm, line width=0.5pt},
	link/.style={draw, line width=1pt},
]
\small

\newcommand\lchip[2]{
	\node[chip,align=center] (chip) at (#1) {Clos\\chip #2};
	\node[io,anchor=west,xshift=-0.2mm] (io) at (chip.east) {};
	\node[fit=(chip) (io),inner sep=0] (chip-#2) {};
}

\newcommand\rchip[2]{
	\node[chip,align=center] (chip) at (#1) {Clos\\chip #2};
	\node[io,anchor=east,xshift=0.2mm] (io) at (chip.west) {};
	\node[fit=(chip) (io),inner sep=0] (chip-#2) {};
}

\lchip{0,0}{1}
\lchip{0,-1.08*\chipsize}{2}

\rchip{\chipsize+\wirewidth,0}{3}
\rchip{\chipsize+\wirewidth,-1.08*\chipsize}{4}

\foreach \x in {1,...,6} {
	\coordinate (tl) at (chip-1.north east);
	\coordinate (tr) at (chip-3.north west);
	\coordinate (bl) at (chip-2.south east);
	\coordinate (br) at (chip-4.south west);
	
	\pgfmathsetmacro\f{\x/7}
	\coordinate (top\x) at ($(tl)!\f!(tr)$);
	\coordinate (bottom\x) at ($(bl)!\f!(br)$);
	\draw[link] (top\x) -- (bottom\x);
}

\begin{scope}[on background layer]
\node[fit=(chip-1) (chip-2) (chip-3) (chip-4), package] {};
\end{scope}

\foreach \y in {1,2,3,4} {
	\foreach \x in{1,2} {
		\pgfmathsetmacro\f{\y/5}
		\coordinate (chip\x-io\y) at ($(chip-\x.north east)!\f!(chip-\x.south east)-(0.15,0)$);
	}
	\foreach \x in{3,4} {
		\pgfmathsetmacro\f{\y/5}
		\coordinate (chip\x-io\y) at ($(chip-\x.north west)!\f!(chip-\x.south west)+(0.15,0)$);
	}
}


\draw[link] (chip1-io1) circle (0.5pt) -- (chip1-io1 -| top1) circle (0.5pt);
\draw[link] (chip2-io1) circle (0.5pt) -- (chip2-io1 -| top1) circle (0.5pt);

\draw[link] (chip3-io1) circle (0.5pt) -- (chip3-io1 -| top6) circle (0.5pt);
\draw[link] (chip4-io1)circle (0.5pt)  -- (chip4-io1 -| top6) circle (0.5pt);

\draw[link] (chip1-io2) circle (0.5pt) -- (chip1-io2 -| top2) circle (0.5pt);
\draw[link] (chip3-io2) circle (0.5pt) -- (chip3-io2 -| top2) circle (0.5pt);

\draw[link] (chip2-io2) circle (0.5pt) -- (chip2-io2 -| top3) circle (0.5pt);
\draw[link] (chip4-io2) circle (0.5pt) -- (chip4-io2 -| top3) circle (0.5pt);

\draw[link] (chip1-io3) circle (0.5pt) -- (chip1-io3 -| top4) circle (0.5pt);
\draw[link] (chip4-io3) circle (0.5pt) -- (chip4-io3 -| top4) circle (0.5pt);

\draw[link] (chip2-io4) circle (0.5pt) -- (chip2-io4 -| top5) circle (0.5pt);
\draw[link] (chip3-io4) circle (0.5pt) -- (chip3-io4 -| top5) circle (0.5pt);

\end{tikzpicture}%

  \label{fig:clos-chip-layout}}
\qquad
\subfloat[mesh layout]{
  
\def\chipsize{18mm}
\def\wirewidth{20mm}

\begin{tikzpicture}[
	chip/.style={draw,align=center,minimum size=0.9*\chipsize,fill=white,shaded},
	io/.style={draw,minimum size=\chipsize,shadedDark},
	package/.style={, draw, inner sep=2mm, line width=0.5pt},
]
\small

\newcommand\chip[2]{
	\node[io] (io) at (#1) {};
	\node[chip] (chip) at (#1) {Mesh\\chip #2};
	\node[fit=(chip) (io),inner sep=0] (chip-#2) {};
}

\chip{0,0}{1}
\chip{1.08*\chipsize,0}{3}
\chip{0,-1.08*\chipsize}{2}
\chip{1.08*\chipsize,-1.08*\chipsize}{4}

\begin{scope}[on background layer]
\node[fit=(chip-1) (chip-2) (chip-3) (chip-4), package] {};
\end{scope}

\end{tikzpicture}%

  \label{fig:mesh-chip-layout}}

\caption{Chip layouts on the silicon interposer.
For a folded Clos network, the chips are positioned beside a wiring channel that contains
a common wire for every connection between two chips. A connection is made by
connecting horizontal wires from each chip to the common wire; diagram
\subfig{clos-chip-layout} shows a wiring pattern for connections from each
chip to every other chip.
For the 2D mesh layout, in \subfig{mesh-chip-layout}, chips are arranged in a
grid and the network is extended directly between chips.}

\label{fig:interposer-layouts}
\end{figure*}

\section{Implementation technology parameters}
\label{sec:parameters}

This section presents and explains the parameters used to characterise current
production technology for the implementation model.
The International Technology Roadmap for Semiconductors (ITRS), which is a set
of documents produced annually by a broad collection of industry members,
provides many of these; others are taken from elsewhere in the literature or
estimated.
Although each parameter is given a specific value, the abstract nature of
the implementation model means it is relatively robust to variations in them.

The remainder of this section is divided into the parameters of the processing
and interposer chips, the main VLSI components and the memories.
\tab{chip-parameters} summarises the parameters for the processing chip and
\tab{interposer-parameters} summarises the parameters for the interposer.

\begin{table*}[t]
\small
\centering
\begin{tabularx}{\textwidth}{p{3.8cm}x{2.5cm}X}
{\bold Parameter} & {\bold Value} & {\bold Note}\tn
\midrule
Process geometry & 28~nm & \tn
FO4 delay & 11~ps & See~\sect{chip-parameters}.\tn
Economical chip sizes & 80-140~mm$^2$ &
  Cost-performance processor~\cite[Tab.~ORTC-2C]{ITRSExec12}.\tn
Metal layers & 8 & M1 logic; M2, 7 \& 8 power \& clock; M3-M6 wiring \tn
Interconnect wire pitch & 125~nm &
  For the global interconnect \cite[Tab.~INTC6 (`12)]{ITRSInterconnect}.\tn
Repeated wire delay & 155~ps/mm & See~\sect{wire-parameters}.\tn
Processor area & 0.10~mm$^2$ & See~\sect{component-areas} \tn
Switch area & 0.05~mm$^2$ & See~\sect{component-areas}.\tn
I/O pad dimensions & 45 $\times$ 225~$\mu$m & See~\sect{chip-parameters}.\tn
Wires per link & 18 & 1 control, 8 data per direction, see~\sect{wire-parameters}.\tn
Power and ground I/Os & 40\% & See~\cite[Tab.~ORTC-4]{ITRSExec12}.\tn
Clock rate & 1~GHz & See~\sect{chip-parameters}.\tn
\end{tabularx}

\caption{Implementation parameters for the processing chip model.}
\label{tab:chip-parameters}

\bigskip
\begin{tabularx}{\textwidth}{p{3.8cm}x{2.8cm}X}
{\bold Parameter} & {\bold Value} & {\bold Note}\tn
\midrule
Process geometry & 65~nm & $\dagger$\tn
FO4 delay & 24~ps & See~\sect{chip-parameters}.\tn
Metal layers & 4 & $\dagger$, M1 \& M2 power \& ground; M3 \& M4 wiring.\tn
Interconnect wire pitch & 2~$\mu$m & $\dagger$, 333 half-shielded wires/mm.\tn
Repeated wire delay & 89~ps/mm & See~\sect{wire-parameters}.\tn
Microbump pitch & 45~$\mu$m & $\dagger$, 493.83~bumps/mm$^2$.\tn
TSV pitch & 210~$\mu$m & $\dagger$, 22~TSVs/mm$^2$.\tn
C4 bump pitch & 210~$\mu$m & $\dagger$, 22~bumps/mm$^2$.\tn
Wires per link & 10 & 1 control, 4 data per direction, see~\sect{wire-parameters}.\tn
\end{tabularx}

\caption{Implementation parameters for the interposer model.
  $\dagger$ parameters based on Xilinx Virtex 7 FPGA
  package~\cite{XilinxWorkshop11}, which integrates four 28~nm FPGA slices on a
  775~mm$^2$ interposer.}

\label{tab:interposer-parameters}
\end{table*}

\subsubsection{Processing chip and interposer}
\label{sec:chip-parameters}

The parameters for the processing chip are based on a 28~nm logic process.
Although 20~nm and 16~nm are also commercially available, they are relatively
new and accordingly less mature, making them substantially more expensive.
28~nm can therefore be considered representative of current devices. The range
of economical chip sizes, 80-140mm$^2$ is based on the ITRS data for a
`cost-performance microprocessor at
production'~\cite[Tab.~ORTC-2C]{ITRSExec12}.

The chip I/O pad area includes the contact and driver circuitry and its
dimensions are estimated based on a 1-to-4 ratio between width and height,
which is characteristic of conventional designs~\cite{Weste05}. The width is
taken to be the pitch of the interposer contacts (microbumps).
An on-chip processor and interconnect clock rate of 1~GHz is chosen as this is
typical of a high-end embedded processor.
The parameters for the silicon interposer and packaging are based on the
passive interposer of the Xilinx Virtex 7 FPGA~\cite{Jones10, Xilinx11,
XilinxWorkshop11}, however it is assumed that wires can be repeated.

All links between switch components on the processing chip have 9 wires in
each direction, allowing up to one byte to be transmitted per cycle with an
additional bit to signal control tokens.
All off-chip links have 5 wires in each direction, allowing up to one byte
to be transmitted every two cycles. The reduction in off-chip link wires is to
reduce the off-chip I/O requirements and interposer wiring.

The delay of an optimally-repeated wire can be estimated with the equation
$$\tau = 1.47\sqrt{\text{FO4}\cdot \hat{R}\hat{C}}$$
where $\tau$ is the delay in picoseconds per unit length; $\hat{R}\hat{C}$ is a
time constant equal to the product of the resistance per unit length and the
capacitance per unit length; and FO4 (fanout-of-4) delay is a constant
related to the particular process technology, defined as the delay of an
inverter, driving four identical copies of itself (see~\cite[\S2.2]{Ho03b} for
more details).
This approximation is based on the derivation by Bakoblu~\cite{Bakoglu90} and
follows the explanation by Ho~\cite[\S4.2]{Ho03b}.
The FO4 delay is estimated for a particular process geometry using the
heuristic $\text{FO4} = 360 \cdot f$, where $f$ is the feature size in $\mu$m,
producing a delay in ps~\cite{Ho01}.

The ITRS provide figures for wire $RC$ delay, which are summarised for a range
of process geometries in \tab{wires}; their method of calculation for the $RC$
delay is described in~\cite[\S7.3 (`12)]{ITRSInterconnect}.
The M1 half pitches of 68~nm and 26.76~nm are the closest matching geometries
of the processing chip and interposer respectively and the data for them are
taken to estimate wire delay.
Using these figures and the above formula for wire delay yields 155~ps/mm for
the 28~nm processing chip and 89~ps/mm for the 65~nm interposer chip.

\label{sec:wire-parameters}

\begin{table*}[t]
\small
\centering
\begin{tabular}{x{2.2cm}x{3cm}cc}
{\bold Process geometry (M1~$\frac{1}{2}$~pitch)} &
{\bold Minimum global wire pitch (nm)} &
{\bold $RC$ delay (ps/mm)} &
{\bold ITRS edition} \\
\midrule
150    & 670 & not given  & 2001\\
90     & 300 & 96         & 2005\\
68*    & 210 & 168        & 2007\\
45     & 154 & 385        & 2010\\
37.84  & 114 & 621        & 2011\\
26.76* & 81  & 1,115      & 2012\\
\end{tabular}

\caption{ITRS data for global wires~\cite{ITRSInterconnect}. The rows marked
with a * are the ones used to determine the wire delay for the processing chip
and interposer.}

\label{tab:wires}
\end{table*}

\subsubsection{Processor and switch components}
\label{sec:component-areas}

The area $A$ of a component at a process geometry $g$ is estimated for a
different geometry $h$ by applying a linear scaling $A_h = A_g/(g/h)^2$, where
$g\geq h$.
The area of the XMOS XCore processor on a 90~nm process is conservatively
estimated to be 1~mm$^2$; on a 28~nm process it is estimated to be 0.10~mm$^2$.
The area of the C104 switch on a 1~$\mu$m process was approximately 40~mm$^2$;
on a 28~nm process it is estimated to be 0.03~mm$^2$.

These figures are consistent with the ARM Cortex-M0
processor~\cite{ARMCortexM0} and the 32$\times$32 SWIFT switch~\cite{Mudge11}.
The Cortex-M0 is a simpler processor than the XCore, having only a 3-stage
pipeline and support for a single hardware thread. (The XCore has a 4-stage
pipeline and eight hardware threads.) On a 40~nm process the M0 has an area of
0.01~mm$^2$ and on 28~nm an estimated area of 0.003~mm$^2$.
The SWIFT switch has an area of 0.35~mm$^2$ on a 65~nm process and an estimated
area of 0.06~mm$^2$ on 28~nm.

\subsubsection{Memory}

\begin{table*}
\small
\centering
\begin{minipage}[t]{\textwidth}
\begin{tabular}{lx{1.4cm}x{1.4cm}x{1.6cm}x{1.9cm}x{1.9cm}x{1.2cm}}
{\bold Type} & {\bold Typical capacity (MB)} & {\bold Cell area factor ($F^2$)}
  & {\bold Area efficiency} & {\bold Current process geometry (nm)}
  & {\bold Density (KB/mm$^2$)} & {\bold Cycle time (ns)}\tn
\midrule
SRAM & $<$8 & 140 & 70\% & 28 & 778.51 & 0.5\tn
eDRAM & 1 - 64 & 50 & 60\% & 28 & 1,868.42 & 1.3\tn
Comm.\ DRAM & $>$64 & 6 & 60\% & 40%
\footnote{High-volume commodity DRAM, in general, lags commodity logic
  by several process technologies.}
& 7,629.39 & 30%
\footnote{Random cycle time ($t_{RC}$) from a 1~Gb Micron DDR3
  device~\cite{MicronDDR3}, contemporary to 28~nm logic.}\tn
\end{tabular}

\caption{Comparison of contemporary memory technologies, with figures from the
2012 ITRS report~\cite[Tab.~SYSD3b]{ITRSSystemDrivers12}. The area factor is
a multiple of square half pitch units, the number given by the process
geometry. The area efficiency is the proportion of area in a memory array
that is occupied by storage cells.}

\label{tab:memory-comparison}
\end{minipage}
\end{table*}

The area and access latency of memory for the processing tiles is estimated using
data from the ITRS~\cite[Tab.~SYSD3b]{ITRSSystemDrivers12}.
\tab{memory-comparison} gives the key characteristics of the dominant forms of
memory technology. Figures for SRAM and eDRAM are given for 28~nm, in line with the
processing chip; 40~nm DRAM is included as a baseline, which is characteristic
of contemporary commodity devices.

Although density, and in particular, cycle time (the time to perform a random
access) will vary with capacity and process technology, typically, memory
capacity increases with process technology and these remain roughly constant
between generations.
In general, eDRAM is 2 to 3 times the density of SRAM and 4 to 5 times less
dense than commodity DRAM. SRAM cycle time is around three times faster than
eDRAM and commodity DRAM cycle time is much higher because of the specialised
process on which it is produced.

Although eDRAM can, in principle, be integrated with microprocessor logic,
there are relatively few devices that use it as a technology, and the
additional process steps that required significantly increase manufacturing
costs. For these reasons, only SRAM technology is considered in the
implementation models.
The tile capacities of 64~KB, 128~KB, 256~KB and 512~KB are selected since they
have a similar area to the processor (0.08~mm$^2$ to 0.66~mm$^2$).

\subsection{Cost and scaling}
\label{sec:scaling}
\label{sec:area-and-wire-delay}

The figures presented in this section are produced from the calculation of
specific instances of the chip layouts for the chosen technology parameters.

\subsubsection{Of the processing chip}

\begin{figure}
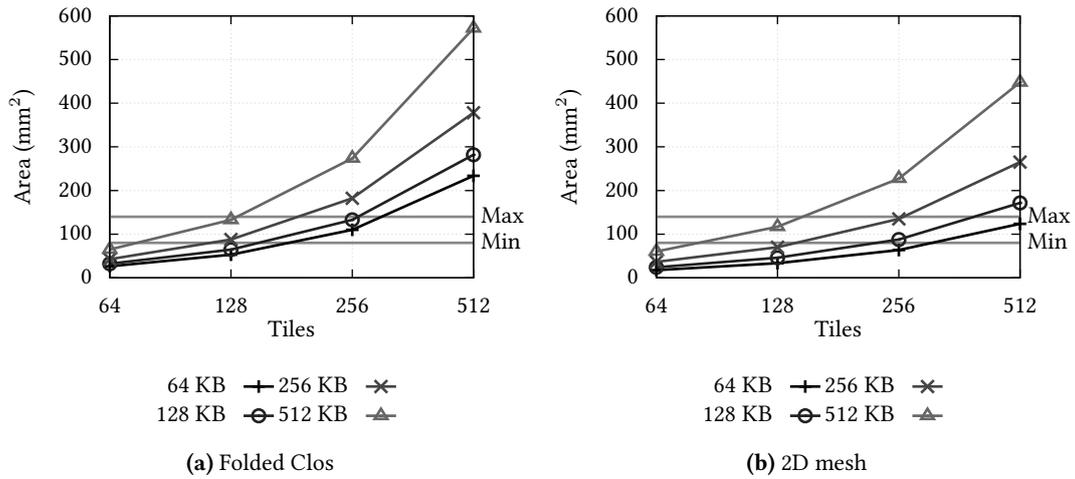

\centering
\subfloat[Folded Clos] {
{\footnotesize\input{results/model/clos-totals-sram}}
}
\subfloat[2D mesh] {
  {\footnotesize\input{results/model/mesh-totals-sram}}
}

\caption{Log-linear plots of the total chip area as a function of the numbers
of tiles.  The grey horizontal lines indicate the range of economical chip
sizes, from 80~mm$^2$ to 140~mm$^2$.}

\label{fig:chip-areas}
\end{figure}

\fig{chip-areas} shows how the total chip area scales with the number of tiles
integrated.  Horizontal lines are drawn to mark the range of economical chip
sizes (80~mm$^2$ to 140~mm$^2$) and marked `Min' and `Max'.
Of the chips that fall in the economical range, the folded-Clos chip requires
13\% to 43\% more area than the interconnect in the 2D mesh chip with the same
number of tiles and memory capacity.
For example, the largest folded-Clos chip with 256 tiles with 128~KB of memory
occupies 132.9~mm$^2$ (of which 44.6~mm$^2$ is occupied by I/O) and the
corresponding 2D mesh occupies 87.9~mm$^2$.

For the folded-Clos chips, apart from the configuration with 128 tiles and
512~KB of SRAM, the wires connecting the tiles to the stage-one switch are less
than 5.5~mm long, have sub-nanosecond delays and thus are single cycle.  All
other delays on wires up to 11.2~mm are less than two nanoseconds and thus have
a two-cycle latency.
For the 2D-mesh chips, wire lengths between switches vary from 1.7~mm to 3.5~mm
with sub-nanosecond single-cycle delays.

\subsubsection{Of the components of the processing chip}

\begin{figure*}
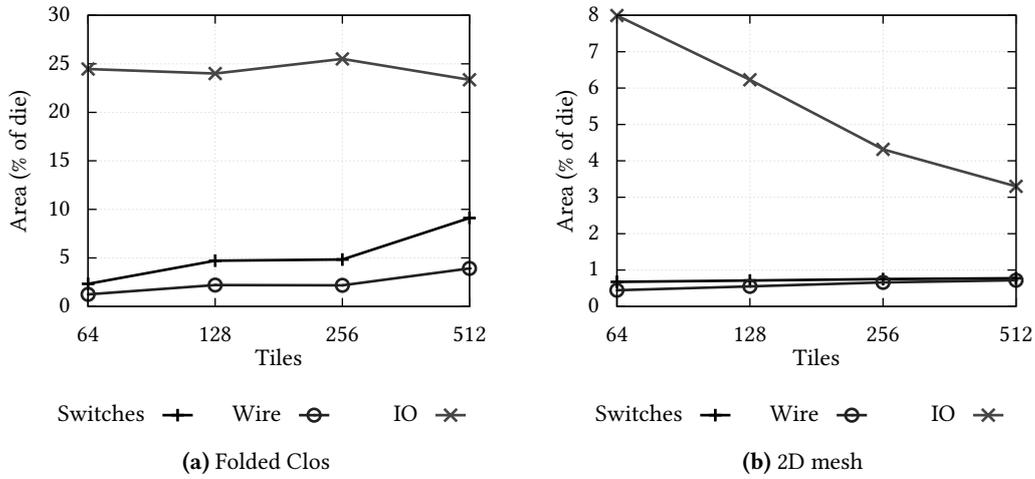

\centering
\subfloat[Folded Clos]{
  {\footnotesize\input{results/model/clos-components}}
  \label{fig:clos-components-per-tile}
}
\subfloat[2D mesh]{
  {\footnotesize\input{results/model/mesh-components}}
  \label{fig:mesh-components-per-tile}
}

\caption{
Log-linear plots of the total area of the switches and wiring components as a
percentage of the total die area for chips with 256~KB of tile memory.
Switch area is calculated as the sum of the switch group area in the folded
Clos. This adds an overhead to the otherwise logarithmic scaling of switch and
wire area.
Both components remain constant in the 2D mesh, apart from a small convergent
growth in the wire area due to a decreasing ratio of switches on the edges of
the 2D mesh to those in the middle.}

\label{fig:chip-component-areas}
\end{figure*}

\fig{chip-component-areas} shows the total area of the switch, wire and I/O
components of the chip as a percentage of the total area for 256~KB tile
memories. These plots show how the component areas scale with respect to the
number of tiles they connect, highlighting the difference between the folded
Clos and 2D mesh in the rate that resources are invested with increasing size.

The total switch area for the folded Clos chips is calculated as the total area
occupied by the switch groups.  Although the number of switches and links
per-tile is proportional to the number of stages and the growth of additional
components is logarithmic in the number of tiles, the area grows more quickly
than this due to the increasing inefficiency of larger switch groups.
The I/O remains relatively constant per tile but occupies a large region of the
die. For the smallest 64~KB memories it occupies approximately 40\% of the
die.
Overall, for the economical chip sizes, the interconnect occupies between 5\%
and 8\% of the die area.

For the 2D mesh, switch area remains constant per tile and the wire grows
slowly because of the decreasing ratio of switches on the edges of the mesh to
those in the middle.
Similarly, the proportion of I/O diminishes as the number of tiles increases
since the number of external connections diminishes relative to the number of
tiles.
Overall, for the economical chip sizes, the 2D mesh interconnect occupies 2\%
to 3\% of the die area.

\subsubsection{Of the interposer}

For a folded Clos system with multiple chips integrated on a silicon interposer,
the percentage area occupied by the wiring channel varies between 2\% for two
128-tile chips with 64~KB memory per tile (234~mm$^2$ total), and up to 42\%
for 16 512-tile chips with 128~KB of memory per tile (1,979~mm$^2$ total).
Accordingly, the minimum and maximum wire delays range from 1~ns to 8~ns, which
correspond to the width and the width plus the height of the channel.
For the 2D mesh, the wire delay is a constant 0.09~ns for all systems since the
chips are tiled and the distance between adjacent pads is constant.

\begin{figure*}
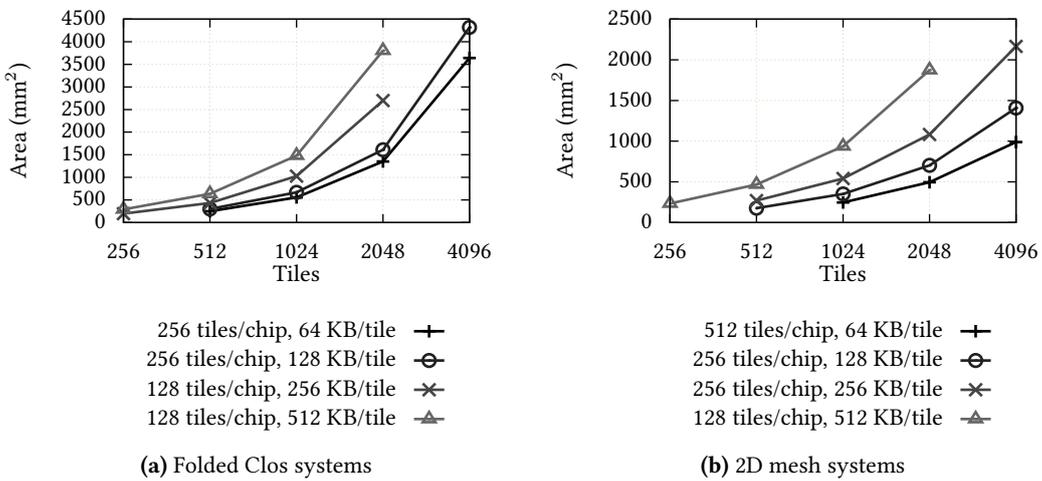

\centering
\subfloat[Folded Clos systems]{
  {\footnotesize\input{results/model/clos-packages-sram}}
  \label{fig:clos-packages-sram}
}
\subfloat[2D mesh systems]{
  {\footnotesize\input{results/model/mesh-packages-sram}}
  \label{fig:mesh-packages-sram}
}

\caption{Log-linear plots of the total area of the interposer for different
configurations of selected economically-sized processing chips.}

\label{fig:package-areas}
\end{figure*}

The plots in \fig{package-areas} summarise the total interposer area for a
range of different system configurations, with varying numbers of different
processing chips.

\clearpage
\section{Experimental methodology}
\label{sec:methodology}

This section presents the methodology by which the performance of the emulation
of a large memory with the proposed parallel architecture is compared to that
of a sequential machine.

\subsection{Sequential machine model}

\def\dramlatencyfootnote{
There is an overhead in accessing a particular row of a DRAM array and
successive accesses to the same row exhibit lower latency.  Consequently,
measurement of DRAM latency is based on two main components.  The column
address strobe ($t_{CL}$) latency, which is the time between specifying a
column address and receiving the data in response, given that the row being
accessed is already open.
The row cycle time ($t_{RC}$), which is the minimum time between the activation
of one row and another; there is a minimum period a row must be active for to
perform a refresh and a non-active row must be precharged before it can be read
from.}

\def\rankfootnote{
A DRAM rank is a set of one or more DRAM chips that are accessed simultaneously
and provide a particular data width.  For standard DRAMs, a channel typically
has a 64-bit data bus and with error-correcting code (ECC) memory, this is
expanded to 72-bits.}

The performance of a sequential machine is used to provide a baseline for the
memory emulation. Performance is presented principally by the relative slowdown
of the emulation compared to the sequential machine executing the same
program.
The modelled performance is based on latency since it is the most difficult
aspect to scale; schemes for scaling bandwidth generally involve replication
and hence may in principle be applied to both systems.

The sequential machine is modelled as a single processor of the same class as
the parallel machine connected to a DRAM system.  A cache is not modelled
directly but memory accesses to areas stored in local memory in the parallel
emulation incur only a single-cycle latency.  The effect of this is similar to
providing the sequential system with a fast cache memory with an 80\% to 90\%
hit rate since global memory accesses constitute between 10\% to 20\% of
executed instructions in the benchmarks. The comparison therefore directly
compares DRAM accesses with emulated accesses. Indeed, any caching scheme for
the sequential machine could equally be implemented as part of the the parallel
emulation.
The clock frequency of both systems is held at 1~GHz and it is assumed the
sequential DRAM can operate at this speed (typical DDR3, which is contemporary
to 28~nm logic, operates well in excess of this).

The access latency is estimated for a modern DRAM by simulation with
DRAMSim2~\cite{Jacob11}.  Performance is measured by performing read and
write accesses to addresses chosen uniformly at random over the address range
and the fixed latency is calculated as the average of these
accesses.\footnote{\dramlatencyfootnote}
Latency measurements are based on random accesses and accesses are issued only
once the last has completed to restrict the memory controller to processing a
single transaction at a time.
Although random access is pessimistic because successive accesses to the same
DRAM page exhibit less latency and schemes such as prefetching do also reduce
latency, these are not considered in the model because they do in principle
apply to both systems.
For a system with 1~Gbit DDR3 chips~\cite{MicronDDR3}, average random-access
latency is measured at 35~ns for a single rank\footnote{\rankfootnote}
with a 1~GB capacity.  For multi-rank system with 2~GB to 16~GB capacities, this
increases to 36~ns due to a small overhead in switching between ranks.  This
choice is unimportant since the empirical analysis is concerned with
obtaining results to within small factors and to demonstrate scaling behaviour.

Typically, DRAMs are packaged in DIMM cards but production processes are moving
towards stacked packages integrated with through-silicon vias and it is
expected that stacks with multiple DRAM chips will be available soon.
Since DRAMSim2 does not model a particular packaging or wire delay, the above
sequential machine model's performance is inclusive of a stacked DRAM
configuration.

\subsection{Benchmarks}
\label{sec:instruction-mix}

The performance analysis is based on two benchmarks: synthetic instruction
sequences and a simple compiler.

\begin{figure}
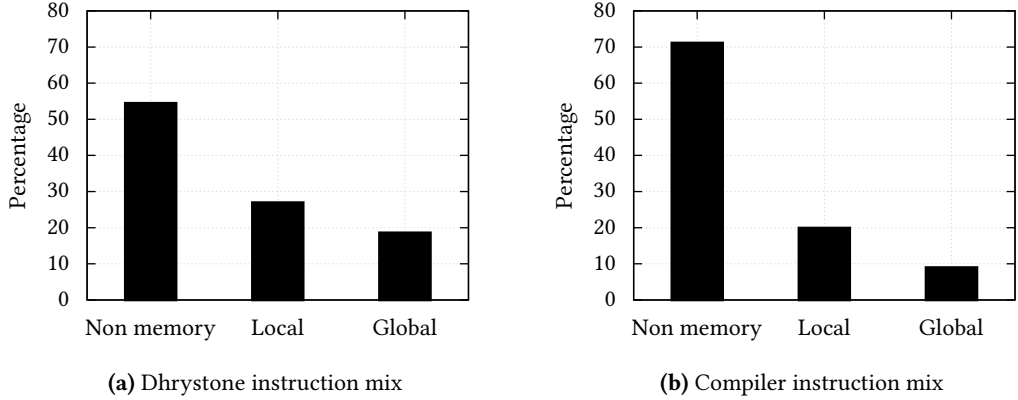

\centering
\subfloat[Dhrystone instruction mix]{
  {\footnotesize\input{results/emulation/mix-dhrystone}}
  \label{fig:mix-dhrystone}}
\subfloat[Compiler instruction mix]{
  {\footnotesize\input{results/emulation/mix-xcmp}}
  \label{fig:mix-xcmp}}

\caption{Instruction mix of the Dhrystone and compiler benchmarks, showing the
proportions of executed non-memory, local memory and global memory
instructions.}

\label{fig:instruction-mix}
\end{figure}

The synthetic instruction sequences contain a particular ratio of global memory
accesses to local memory and non-memory operations, to coarsely characterise
sequential programs.
Non-memory instructions are, for example, arithmetic and branching;
local-memory instructions include all accesses to the program, stack and
constant data; and global memory instructions access static data and the heap
region.
The ratios are chosen based on the instruction mix of the Dhrystone
benchmark~\cite{Weicker84Dhrystone} (which is intended to characterise integer
general-purpose sequential programs) and at points over the range of potential
different ratios to demonstrate the effect they have on performance.
\fig{mix-dhrystone} shows the proportions of different types of instructions
executed for the benchmark. 
In all experiments, the proportion of local memory accesses in the synthetic
sequences is fixed at 20\%.

The compiler benchmark provides an example of a realistic general-purpose
sequential application and a means of compiling sequential programs to the
parallel architecture.
A modified version of the compiler emits message-passing sequences in the place
of global memory accesses in order to generate sequential programs that
interact with an emulated global memory (the effect on code size is discussed
in~\sect{code-size}).
\fig{mix-xcmp} shows the proportions of different types of instructions
executed for the compiler benchmark.
%

\subsection{Performance model}
\label{sec:performance-model}

\def\ttile{t_{\text{tile}}}
\def\tswitch{t_{\text{switch}}}
\def\topen{t_{\text{open}}}
\def\ccontention{c_{\text{cont}}}
\def\tlink{t_{\text{link}}}
\def\tserial{t_{\text{serial}}}
\def\tserialinter{t_{\text{serial-inter}}}
\def\tserialintra{t_{\text{serial-intra}}}
\def\tdelay{t_{\text{delay}}}

The system model presented in \sect{implementation} provides values for the
latency of signal transmissions along links and their bandwidth. The
performance of communication depends on some additional parameters related to
the latency of the switching elements in the network.
The following describe these parameters and formulate a performance model for
the folded Clos and 2D mesh networks.

Given an interconnection network represented by a graph $G=(V,E)$ whose
vertices $V$ represent nodes containing a switch and zero or more tiles, and
edges $E$ that represent communication links, the latency of a message sent
from processor $s\in V$ to processor $t\in V$ depends on:
\bit
\item $\ttile$, the latency of the link between the tile and the switch;
\item $\tswitch$, the switch latency;
\item $\topen$, the additional latency to open a route through the switch;
\item $\ccontention$, the switch contention factor;
\item $d(s, t)$, the length of the path, $p$, in the network ($d(s,t)=|p(s,t)|$);
\item $\tlink(u, v)$, the latency of the link between nodes $u,v\in V$;
\item $\tserial$, the serialisation latency, which is determined by the message
  length and channel bandwidth, such that if $s$ and $t$ are on the same chip,
  then $\tserial=\tserialintra$ and if they are on different chips, then
  $\tserial=\tserialinter$.
\eit
When the route between $s$ and $t$ is not already open, message latency is
calculated as
\begin{align*}
t_{\text{closed}}(s,t) =& 2\ttile + \tserial +
(d(s,t)+1)(\topen+\tswitch\cdot\ccontention)\\*
& + \sum_{\ell \in p(s,d)}\tlink(\ell)
\end{align*}
and when the route is open it is calculated as
\begin{align*}
  t_{\text{open}}(s,t) = & 2\ttile + \tserial +
  (d(s,t)+1)\cdot\tswitch\cdot\ccontention\\*
& + \sum_{\ell \in p(s,d)} \tlink(\ell)
\end{align*}
Both models are comprised of the sum of four latency components:
tile-to-switch, serialisation, switch traversal and link traversal.
With shortest-path routing, $d(s,t)$ is the minimum distance between $s$ and $t$.

\def\xmpfootnote{
The device was the XMP-64~\cite{XMP64}, an array of 16 quad-core
XS1-G4 chips~\cite{XS1-G4-512Datasheet}.  For a general discussion
of the methodology for taking performance measurements and other specific
results with the XMP-64, the reader is referred to~\cite{Hanlon09}.}

Values for the parameters of the above latency are summarised in
\tab{latency-parameters}.
The estimates are used to calculate the link and serialisation latencies and
the switch latencies are estimated by fitting measurements taken with an XMOS
device to the above latency model.\footnote{\xmpfootnote}
Measurements taken for the other parameters are included for comparison, noting
that there is no serialisation latency on chip because the tiles
are connected to the switch with 8-bit links; off-chip, the link bandwidth
is one byte every four cycles.

\begin{table}
\centering
\begin{tabular}{lcx{1.5cm}x{1.5cm}}
{\bold Parameter} &
{\bold Symbol} &
{\bold Value (cycles)} &
{\bold XMP-64 (cycles)}\tn
\midrule
Tile-to-switch latency    & $\ttile$    & see~\sect{area-and-wire-delay} & 1\tn
Switch latency            & $\tswitch$  & 2 & 2\tn
Latency to open a route   & $\topen$    & 5 & 5\tn
Ser.\ latency intra-chip  & $\tserialintra$  & 0 & 0\tn
Ser.\ latency inter-chip  & $\tserialinter$  & 2 & 4\tn
Link latency              & $\tlink$ & see~\sect{area-and-wire-delay}
& 2 on-chip, 3 off-chip\tn
\end{tabular}

\caption{Parameters for the network performance model. The switch latencies are
based on measurements made with the XMP-64; the other measurements are included
for comparison.}

\label{tab:latency-parameters}
\end{table}

\section{Performance results}
\label{sec:results}

In this section, the results of experiments with the modelled systems are
presented. Absolute latency is considered first since this determines the
performance of the emulated memory and is the most difficult characteristic to
scale, after this, the performance of the benchmarks are presented.

\subsection{Absolute latency}

\begin{figure}
\centering
{\footnotesize\input{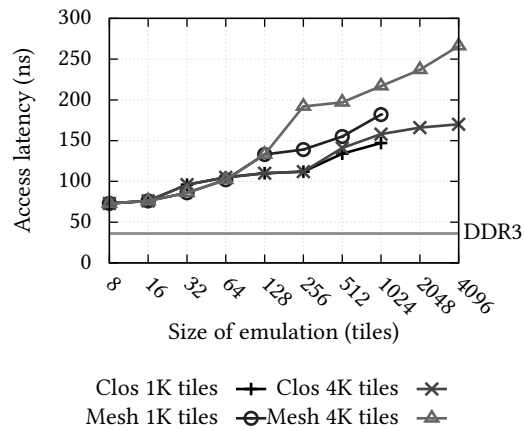}}
\label{fig:access-latency-sp}

\caption{Log-linear plots showing how the absolute memory latency scales for
1,024- and 4,096-tile systems as the number of tiles in the emulation
increases.}

\label{fig:access-latency}
\end{figure}

\fig{access-latency} shows how the average access latency of random reads and
writes in the emulated memory scales as the number of tiles is increased in the
emulation. The baseline latency measured from the simulated DDR3 memory is
included for comparison.

%
Performance of the folded Clos network clearly reflects the logarithmic growth
of the network diameter and the latency incurred by additional stage in systems
larger than 256 tiles can be seen.
Latency in the 2D mesh increases linearly with the size of the emulation, with
a change of gradient as communications occur between chips.
Overall, the folded Clos delivers access latency that is within a factor of
approximately 2 to 5, relative to a sequential machine with a DDR3 memory. The
performance of the two networks is similar on-chip but the 2D mesh incurs a
30\% to 40\% overhead relative to the Clos for larger multi-chip emulations.

It is important to note that the results for absolute latency are based on the
sequential and parallel systems having the same clock speed.
An increase in clock speed for the sequential system relative to the parallel
one would only improve bandwidth because the inherent latency of a DRAM cannot
be improved. However an increase in clock speed for the parallel system would
improve latency because the network would operate faster.

\subsection{Benchmark performance}

\fig{benchmark-slowdowns} shows the performance of the synthetic Dhrystone and
compiler benchmarks under a range of different system parameters.
The general behaviour reflects that of \fig{access-latency}.  The absolute
access latency of the emulated memory is high, a factor of 3 to 4 times that of
a specialised sequential machine for the folded Clos systems, but the effect of
this is diluted by fast local accesses and other non-memory operations.
The folded Clos systems can deliver an emulation with a slowdown of between
approximately 2 to 3 up to 4,096 tiles over the sequential machine.
The performance of the 2D mesh systems is similar to the folded Clos systems up
to approximately 128 tiles. Beyond this, the performance of the 2D meshes
deteriorate relative to the folded Clos, in which latency only increases
slowly.
With both systems, the execution of Dhyrstone is less efficient due to the
higher proportion of global accesses. It is also interesting to note that up to
16 tiles, there is a speedup over the sequential machine because the tiles are
attached to a single switch.

In general, as the ratio of global memory operations to local and non-memory
operations decreases, the slowdown does also, converging to a worst case of 1.5
to 2.5 overhead. This is the ratio between the sequential machine with the DDR3
memory and the parallel emulation shown in \fig{access-latency}.
This trend can also be seen in \fig{synthetic-ranges}, which shows the
emulation performance for the synthetic benchmark with different proportions of
global memory accesses (between 0\% and 50\%).
For general sequential programs where there is a mix of operations and 10\% to
20\% global accesses.

\begin{figure*}
\centering
\subfloat[Dhrystone]{
  {\footnotesize\input{results/emulation/synthetic-dhrystone-sp}}
  \label{fig:dhrystone-sp-slowdown}
}
\subfloat[Compiler]{
  {\footnotesize\input{results/emulation/xcmp-time-sp}}
  \label{fig:xcmp-sp-slowdown}
}

\caption{Log-linear plots of the performance of the synthetic Dhrystone and
  compiler benchmarks, relative to the sequential machine, using an emulated
  memory on 1,024- and 4,096-tile systems.}

\label{fig:benchmark-slowdowns}
%
\subfloat[Folded Clos]{
  {\footnotesize\input{results/emulation/synthetic-range-clos-sp}}
  \label{fig:synthetic-range-clos-sp}
}
\subfloat[2D mesh]{
  {\footnotesize\input{results/emulation/synthetic-range-mesh-sp}}
  \label{fig:synthetic-range-mesh-sp}
}

\caption{Log-linear plots showing the emulation slowdown, relative to the
  sequential machine, over a range of instruction mixes, with proportions of
  global accesses varying between 0\% to 50\%, for 1,024- and 4,096-tile
  systems. The proportion of local memory access is fixed at 20\%, based on the
Dhrystone and compiler instruction mixes.}

\label{fig:synthetic-ranges}
\end{figure*}

\subsection{Program binary size}
\label{sec:code-size}

Since each memory reference is written as a communication sequence (listed
in~\sect{emulation-scheme}), the size of the program binary increases. For
loads, there is an overhead of two instructions and for stores, there is an
overhead of three.
For the version of the compiler that uses the emulated memory, when it compiles
itself, the size of its executable binary increases by 8\%. This however is a
small price compared to the amount of extra memory that can be provided.

\section{Related work}
\label{sec:related}


The need for scalable general-purpose systems has led to a number of proposals
for single-chip tiled parallel architectures.
Examples include
the MIT Raw~\cite{Raw97} and descendant Tile~\cite{Tile07} architectures,
Smart Memories~\cite{SmartMemories00},
the PicoChip architecture~\cite{PicoChip06},
the UC Davis AsAP architecture~\cite{AsAP08},
the Intel Xeon Phi processor~\cite{Intel13XeonPhi},
the Kalray MPPA~\cite{KalrayMPPA12}
and the Cavium ThunderX~\cite{ThunderX14}.
However, despite an established theory of general-purpose parallel computation
which requires low-diameter high-capacity networks~\cite{Leiserson85,
Valiant90b, Leighton91}, these systems neglect this theory and generally employ
a 2D mesh-based interconnect.
The results in this paper indicate that at zero load, 2D meshes may be suitable
for memory emulations with modest numbers of tiles. However, it will be
difficult to maintain efficiency with parallel workloads because the effects
of congestion will increase latency.

Clos~\cite{Clos53} networks and the related fat-tree~\cite{Leiserson85} network
have been widely studied and used for interconnecting parallel systems.
They are well established as interconnects for large systems comprising
thousands of processors in datacentre- and supercomputer-class systems.
On chip, various proposals for electrical, e.g.~\cite{Balfour06, Ludovici09,
Kao11, Jiang14}, and optical, e.g.~\cite{Joshi09, Zhang10}, Clos/fat-tree
networks have been made in order to reduce latency, improve capacity and simplify
programming. These proposals however only consider relatively small systems up to
64 processors and low-degree 2$\times$2 to 16$\times$16 crossbar switches.

\section{Conclusion}
\label{sec:conclusion}

This paper has proposed a general-purpose parallel computer architecture and a
practical implementation for it using current production technologies. It is
able to efficiently emulate large memories with collections of small-memory
processing tiles, thus supporting the execution of sequential programs with
large memory requirements. The resulting machine is capable of switching
between executing highly-parallel and sequential programs, or allowing
the programmer to compose parallel programs with large-memory sequential
components.

The empirical performance results presented demonstrate that the emulation can
be performed efficiently, with just a factor of 2 to 3 overhead for general
sequential programs, compared to a conventional sequential machine. This small
penalty however can be mitigated by exploiting parallelism in the program.
The cost of the network, the central component of the system, is also
relatively low. An on-chip folded-Clos network occupies approximately 7\% of
the die, and off chip, using silicon interposers to provide high density
wiring, it occupies approximately 30\% of the interposer die.  Overall, this
represents a 20\% to 30\% investment in the interconnect, compared to around
5\% with a 2D mesh network.

The presented performance and implementation-cost results from the high-level
modelling demonstrate what can be expected from the systems examined. This
level of investigation is sufficient to reveal the scaling behaviour and the
approximate constant factors involved.

\bibliographystyle{plain}
\bibliography{refs}

\end{document}